# Insight into the effect of force error on the thermal conductivity from machine-learned potentials

Wenjiang Zhou[1,2†], Nianjie Liang[1†], Xiguang Wu[3], Shiyun Xiong[3*], Zheyong Fan[4*], and Bai Song[1,5,6*]

[1]*Department of Energy and Resources Engineering, Peking University, Beijing 100871, China.*

[2]*School of Advanced Engineering, Great Bay University, Dongguan 523000, China.*

[3]*Guangzhou Key Laboratory of Low-Dimensional Materials and Energy Storage Devices, School of Materials and Energy, Guangdong University of Technology, Guangzhou 510006, China.*

[4]*College of Physical Science and Technology, Bohai University, Jinzhou 121013, China.*

[5]*Department of Advanced Manufacturing and Robotics, Peking University, Beijing 100871, China*

[6]*National Key Laboratory of Advanced MicroNanoManufacture Technology, Beijing, 100871, China*

---

[*]Author to whom correspondence should be addressed; e-mail: syxiong@gdut.edu.cn (S. Xiong), brucenju@gmail.com (Z. Fan), songbai@pku.edu.cn (B. Song)

**Abstract**:

Machine-learned potentials (MLPs) have been extensively used to obtain the lattice thermal conductivity via atomistic simulations. However, the impact of force errors in various MLPs on thermal transport has not been widely recognized and remains to be fully understood. Here, we employ MLP-driven molecular dynamics (MD) and anharmonic lattice dynamics (LD) to systematically investigate how the calculated thermal conductivity varies with the force errors, using boron arsenide as a prototypical material. We consistently observe an underestimation of thermal conductivity in MD simulations with three different MLPs including the neuroevolution potential, deep potential, and moment tensor potential. We provide a robust extrapolation scheme based on controlled force noises via the Langevin thermostat to correct this underestimation. The corrected results achieve a good agreement with previous experimental measurement from 200 K to 600 K. In contrast, the thermal conductivity values from LD calculations with MLPs readily align with the experimental data, which is attributed to the much smaller effects of the force errors on the force-constant calculations.

# I. INTRODUCTION

Thermal conductivity ($\kappa$) is the central physical property that dictates the transfer of heat, especially in solids, with important applications in diverse fields such as electronics thermal management [1, 2] and thermoelectric energy conversion [3-5]. Consequently, the prediction and manipulation of thermal conductivity are subjects of broad interest in condensed matter physics. In non-metals such as intrinsic semiconductors or dielectrics, heat is mainly conducted by the lattice variations or phonons. The lattice thermal conductivity can be computed from molecular dynamics (MD) simulations [6-10] or phonon Boltzmann transport equation (BTE) which is grounded in the theory of lattice dynamics (LD) [9, 11, 12]. Foundational to both approaches is the computation of atomic forces, which is usually achieved by using either empirical interatomic potentials such as the Tersoff [13] and Stillinger–Weber potentials [14], or based on density-functional theory (DFT) calculations. However, these approaches are often hindered either by their limited accuracy or the high computational cost.

Recently, a variety of machine-learned potentials (MLPs) have emerged as a promising approach to address such limitations. These potentials are typically trained on a set of reference data including energies, forces, and virial stresses computed using quantum mechanical techniques for a set of representative atomic configurations. Many of the MLPs have already been extensively applied in modeling thermal transport. By leveraging MLP-driven MD (MLP-MD) and MLP-driven BTE (MLP-BTE) simulations, the thermal conductivity values of various materials have been computed including crystalline silicon (Si) [15, 16], boron arsenide (BAs) [17-19], gallium oxide ($Ga_2O_3$) [20-22], lithium hydride (LiH) [23], lead telluride [24], liquid water [25-27], amorphous Si [28], and glass [29]. Despite these notable achievements, several previous studies have revealed that for materials with relatively high thermal conductivity, such as cobalt antimonide ($CoSb_3$) [30], Si [15, 16], gallium arsenide (GaAs) [15], LiH [23], and graphene [15], the values predicted by MLP-MD simulations are generally lower than the corresponding experimental results. In our previous work [15], this underestimation is attributed to the phonon scatterings caused by the finite force errors intrinsic to the MLPs, and is corrected via a simple linear extrapolation.

In this work, we further investigate the impact of force errors on the thermal conductivity as calculated by methods based on MLPs. We employ both MLP-MD and MLP-LD to investigate thermal transport in BAs, an emerging high-$\kappa$ semiconductor (~1300 Wm$^{-1}$K$^{-1}$) with great technological potentials [31-33]. To begin with, a toy model is trained on a dataset obtained from an empirical Tersoff potential, which leads to a notable underestimation of the thermal conductivity from MLP-MD simulations as compared to the MD results calculated using the original Tersoff potential. Subsequently, three MLP models based on different frameworks (neuroevolution potential, deep potential, and moment tensor potential) are trained based on a dataset from DFT calculations. A similar underestimation of the thermal conductivity from MLP-MD simulations is again observed as compared to previous experimental measurements. To address this discrepancy, a linear extrapolation scheme was proposed in our previous work [15]. Here, we extend this method to achieve more robust predictions by incorporating second-order effects. Our corrected values align closely with the corresponding experimental data from 200 K to 600 K. We also investigate the thermal conductivity of BAs via MLP-LD simulations and the results are in good agreement with the corrected $\kappa$ from MLP-MD.

## II. COMPUTATIONAL METHODS

### A. Molecular dynamics simulation

We employ both the equilibrium molecular dynamics (EMD) and the homogeneous nonequilibrium molecular dynamics (HNEMD) methods to calculate $\kappa$. The thermal conductivity tensor in EMD simulations is determined from the heat current autocorrelation function (HCACF) based on the Green-Kubo formula [6, 7]:

$$\kappa_{\mu v} = \frac{1}{k_B T^2 V} \int_0^\infty <\langle J_\mu(0) J_v(t) \rangle>_e \, dt, \qquad (1)$$

where $<\langle J_\mu(0) \cdot J_v(t) \rangle>_e$ represents the ensemble average of HCACF in equilibrium states, $V$ is the volume, $k_B$ is the Boltzmann constant, $T$ is the absolute temperature, and $t$ is the correlation time. In HNEMD simulations, the thermal conductivity tensor in the linear-response regime is expressed as [8, 24, 34]:

$$\frac{<J_\mu(t)>_{ne}}{TV} = \sum_v \kappa_{\mu v} F_e^v, \qquad (2)$$

where $<J_\mu(t)>_{ne}$ represents the ensemble average of the heat current in non-equilibrium simulations, and $F_e^v$ is a driving force parameter acting on the $v$ direction. The potential part of the heat current contributes primarily to the heat conduction in solids [35]: $\mathbf{J}_{pot} = \sum_i \mathbf{W}_i \cdot \mathbf{v}_i$. The per-atom virial is expressed as: $\mathbf{W}_i = \sum_{j \neq i} \mathbf{r}_{ij} \otimes \frac{\partial U_j}{\partial \mathbf{r}_{ji}}$. Here, $\mathbf{v}_i$ is the velocity of atom $i$, $\mathbf{r}_{ij}$ is position difference defined as $\mathbf{r}_{ij} \equiv \mathbf{r}_j - \mathbf{r}_i$, and $U_j$ is the site potentials of atom $j$. Notably, the current implementation of the heat current in the Large-scale Atomic/Molecular Massively Parallel Simulator (LAMMPS) is generally inaccurate for many-body potentials and one must utilize all the nine components of the per-atom virial [35]. In addition, the widely used moment tensor potential (MTP) method [36] exhibits an incorrect implementation of the heat current [34]. In accordance with energy conservation, the accumulated heat from the atoms should correspond to that from the thermostats, but the existing implementation in the MLIP package does not meet this condition [34]. Here, we fix the issue in a modified version of the MLIP package and make it publicly available [37].

First, we develop a toy MLP model using a dataset calculated from the empirical Tersoff potential. Further, we also train three MLPs including the neuroevolution potential (NEP) [24], deep potential (DP) [38, 39], and MTP [36] based on the *ab initio* dataset from the Vienna Ab initio Simulation Package (VASP) [40, 41]. The preparation of the atomic structures, DFT calculations, and training processes are detailed in the Supplemental Material [42]. Based on both the Tersoff potential and the MLPs, we use the HNEMD method as implemented in Graphics Processing Units Molecular Dynamics (GPUMD) [43] to calculate $\kappa$. A supercell of BAs with the dimensions of 48.2 Å×48.2 Å×48.2 Å containing 8000 atoms is used. The periodic boundary conditions are applied in all dimensions. The time step is chosen as 1 fs. The structure is equilibrated under an isothermal-isobaric (*NPT*) ensemble for 1 ns at zero pressure and the target temperature, followed by another 1 ns in the canonical (*NVT*) ensemble. The heat current is sampled for 10 ns in the *NVT* ensemble. These conditions are generally employed unless otherwise specified. Since the DP and MTP models are not supported in HNEMD, we conduct EMD simulations with the LAMMPS package [44]. In EMD simulations, the structure is first equilibrated for 200 ps in the *NPT* ensemble, followed by another 200 ps in the *NVT* ensemble. The heat current is then sampled for 6 ns under the microcanonical (*NVE*) ensemble with a correlation time of 1.2 ns.

## B. Anharmonic lattice dynamics

In addition to MD simulations, we also solve the linearized phonon BTE to calculate the lattice thermal conductivity as:

$$\kappa_{\mu\nu} = \frac{1}{N_{\mathbf{q}}\Omega k_\mathrm{B} T^2} \sum_{p\mathbf{q}} n_{p\mathbf{q}}^0 (n_{p\mathbf{q}}^0 + 1)(\hbar\omega_{p\mathbf{q}})^2 v_{p\mathbf{q}}^\mu v_{p\mathbf{q}}^\nu \tau_{p\mathbf{q}}, \tag{3}$$

where $N_{\mathbf{q}}$ is the number of grid points, $\Omega$ is the volume of a unit cell, $k_\mathrm{B}$ is the Boltzmann constant, $T$ is the temperature, $\hbar$ is the reduced Planck constant, $n_{p\mathbf{q}}^0$ is the occupation number for mode $p$ with wave vector $\mathbf{q}$, $\omega_{p\mathbf{q}}$ is the phonon frequency, $v_{p\mathbf{q}}^\mu$ is the phonon group velocity along the Cartesian direction $\mu$, and $1/\tau_{p\mathbf{q}}$ is the scattering rate (the inverse of phonon lifetime). The harmonic and anharmonic force constants are computed by using the finite-displacement method as implemented in the Phonopy [45, 46], thirdorder [47], and fourthorder [48] codes. The atomic forces of each perturbed structure are calculated based on our trained MLPs. A supercell containing 4×4×4 unit cells is used. The cutoff radii for the thirdorder and fourthorder force constants are set to the 5th and 2nd nearest neighbors, respectively. Then, the three-phonon and four-phonon scattering rates are calculated with a 17×17×17 q-grid using the modified FourPhonon code [49], which is an extension module of ShengBTE package [47]. These parameters are consistent with those employed in Ref. [50]. To accelerate the four-phonon calculation, we use a sampling-estimation approach with the sampling size of $10^5$ [51].

### III. RESULTS AND DISCUSSTIONS

#### A. Toy model based on Tersoff potential

To begin with, we employ the training dataset obtained from the empirical Tersoff potential to train a NEP model for BAs. The parameters in the Tersoff potential for BAs are obtained from Ref. [52]. Details of the training dataset and the training process are presented in the Supplemental Material [42]. The root-mean-square force error for the NEP model (denoted as $\sigma_{\text{mlp}}$) at 300 K is 21.98 meV/Å. Then, we perform HNEMD simulations to compute the room-temperature thermal conductivity of BAs. Hence, the calculated values from the Tersoff potential serve as an ideal reference which enables us to directly elucidate the impact of force errors in MLP-MD simulations.

In Fig. 1, we display the calculated room-temperature thermal conductivity from the HNEMD simulations based on the trained NEP model and the Tersoff potential. With the Tersoff potential, the calculated $\kappa$ for BAs is 2714±188 Wm$^{-1}$K$^{-1}$, in good agreement with previous result from EMD simulations [53]. In dramatic contrast, the value predicted by the toy NEP model is only 1376±46 Wm$^{-1}$K$^{-1}$. This reduction of $\kappa$ can be attributed to the finite force errors inherent in MLPs. Similar underestimations have also been observed in CoSb$_3$ [30], Si [15, 16], GaAs [15], LiH [23], and graphene [15], as compared to the corresponding experimental measurements.

In our previous work [15], we deliberately introduced varying levels of force noise through the Langevin thermostat (denoted as $\sigma_L$) in the MD simulations. The total force error $\sigma_{\text{total}}$ can be calculated as $\sigma_{\text{total}} = \sqrt{\sigma_{\text{mlp}}^2 + \sigma_L^2}$. Based on a first-order approximation and Matthiessen's rule, $1/\kappa$ should be proportional to $\sigma_{\text{total}}$:

$$\frac{1}{\kappa} = \frac{1}{\kappa_0} + \alpha \sigma_{\text{total}}, \qquad (4)$$

where $\kappa_0$ is the intrinsic thermal conductivity and $\alpha$ is a fitting parameter. One can then obtain $\kappa_0$ by extrapolating to the limit of zero force error using Eq. (4). However, a key challenge is to accurately determine the value of $\sigma_L$ below which Eq. (4) remains valid. In fact, we find that the first-order approximation is not robust for BAs. This is attributed to its exceptionally high thermal conductivity, which leads to a rather narrow range of applicability for Eq. (4). Inspired by the critical role of four-phonon scatterings in BAs and similar materials, we further add a second-order term, leading to the following modified expression:

$$\frac{1}{\kappa} = \frac{1}{\kappa_0} + \alpha \sigma_{\text{total}} + \beta \sigma_{\text{total}}^2. \tag{5}$$

Here, $\beta$ is also a fitting parameter. As shown in Fig. 1, the best-fit curve using Eq. (5) overlaps well with the NEP-HNEMD simulated results, and the extrapolation to $\sigma_{\text{total}} = 0$ closely matches the data point directly calculated from the Tersoff potential. Therefore, Eq. (5) exhibits improved robustness over Eq. (4) and will be applied in further analyses below.

### B. Molecular dynamics with machined-learned potentials

Subsequently, we investigate thermal transport in BAs by training MLPs using an *ab initio* dataset. To fully demonstrate the impact of MLPs, we employ three distinct training frameworks: NEP [24], DP [38], and MTP [36]. Details of the dataset and the training process are provided in the Supplemental Material [42]. NEP-HNEMD simulations as implemented in GPUMD are carried out to obtain the $\kappa$ values for various force errors and temperatures. Since the DP and MTP models are not supported in GPUMD, we perform EMD simulations using LAMMPS to obtain the thermal conductivity. As described in the Computational Methods section, the heat current calculation as implemented in the MLIP/LAMMPS interface is inaccurate for many-body potentials. Here, we address this issue by modifying the MLIP package and thus obtain the correct heat flux [37]. Figures 2(a) and 2(b) plot the accumulated heat in BAs during a non-equilibrium steady state as a function of time, using the original and modified MLIP packages, respectively. It can be observed that the accumulated heat from the atoms diverges from that of the thermostats as time progresses. This difference becomes more pronounced for two-dimensional materials such as graphene, as shown in Fig. S2 and in Fig. 3 of Ref. [34, 54]. In our modified version, the accumulated heat from the atoms displays only minor fluctuations with that from the thermostats [Fig. 2(b) and Fig. S2(b)]. As displayed in Figs. 2(c) and 2(d), the incorrect implementation of heat flux results in a thermal conductivity of BAs that is 18% higher in MTP-MD simulations.

In Fig. 3(a), we show the calculated room-temperature thermal conductivity using all three MLPs with respect to the force errors. At 300 K, the $\sigma_{\mathrm{mlp}}$ values of NEP, DP, and MTP are 15.46 meV/Å, 14.16 meV/Å, and 13.06 meV/Å [42], respectively. The NEP-HNEMD, DP-EMD, and MTP-EMD approaches yield values of approximately 932.8±15.3 Wm$^{-1}$K$^{-1}$, 849.1±96.7 Wm$^{-1}$K$^{-1}$, and 777.5±71.1 Wm$^{-1}$K$^{-1}$, respectively. In addition, the NEP-HNEMD calculations are consistent with that from the NEP-EMD simulations (901.5±75.5 Wm$^{-1}$K$^{-1}$), as also reported by a previous study [55]. Note that phonon-isotope scattering is not included in these MD simulations. For isotopically pure BAs, the thermal conductivity value derived from BTE calculations is 1332 Wm$^{-1}$K$^{-1}$ [56]. In comparison, experimental measurements indicate a thermal conductivity range between 1000 and 1300 Wm$^{-1}$K$^{-1}$ at 300 K [31-33]. Thus, our simulations further suggest that the force errors originating from the MLPs in MLP-MD tend to underestimate the thermal conductivity.

Similarly, the thermal conductivity from MLP-MD simulations decreases as the force error introduced by the Langevin thermostat increases. HNEMD is particularly noteworthy for its computational efficiency, requiring only three to five independent simulations, as illustrated in Fig. S3 [42]. Therefore, we focus on fitting the NEP-HNEMD results and extrapolating to the limit of zero force error to obtain the intrinsic thermal conductivity $\kappa_0$ of BAs. The fitting curve in Fig. 3(a) indicates a strong correlation with the NEP-HNEMD data. In addition, the thermal conductivity values computed from DP-EMD and MTP-EMD approaches align well with our extrapolation curves using Eq. (5), while the linear correction method based on Eq. (4) is inadequate. Using the open DP model (with a $\sigma_{\text{mlp}}$ value of 27.9 meV/Å at 300 K) for BAs trained by Wu *et al*. [57], the predicted thermal conductivity is 690.6±80.6 Wm$^{-1}$K$^{-1}$, which is also in good agreement with our results.

We further perform MLP-MD simulations with various force errors to compute the thermal conductivities at higher temperatures. The $\sigma_{\text{mlp}}$ of the three MLPs are listed in Table. S1 of the Supplemental Material [42]. The NEP-HNEMD results are primarily used to demonstrate the effectiveness and robustness of Eq. (5). For DP and MTP potentials, we only focus on the effect of $\sigma_{\text{mlp}}$ considering their relatively higher computational cost. As plotted in Figs. 3(b) to 3(d), the curves fitted by using Eq. (5) at higher temperatures are in good agreement with the results from NEP-HNEMD, DP-EMD, and MTP-EMD simulations. As the temperature continues to rise, the intrinsic anharmonic phonon-phonon scattering increases, leading to a reduction in the relative contribution of the force errors to higher-order phonon scattering. Therefore, the first-order assumption in the small force error range becomes valid. Equation (4) can then be employed to fit the $\kappa$ values, yielding consistent results as illustrated in Fig. 3(d) at 600 K.

Now that the relation between the thermal conductivity and force errors is established, we can extrapolate to the limit of $\sigma_{\text{total}} = 0$ to obtain $\kappa_0$. In Fig. 4, we display the original and corrected thermal conductivities as a function of temperature. The extrapolated results are in excellent agreement with experimental values from 300 K to 600 K. This demonstrates that our newly proposed method can effectively correct the general underestimation of thermal conductivity in MLP-MD, even within the non-linear force error region. Here, the careful selection of $\sigma_L$ to ensure the applicability of linear fitting is no longer required. To further demonstrate its versatility and robustness, we apply Eq. (5) to the case of GaAs discussed in our previous work [15]. As displayed in Fig. S4 of the Supplemental Material [42], the good fitting and extrapolation results again confirms the effectiveness of our approach, even across different material systems.

### C. Lattice dynamics via machined-learned potential

Finally, we evaluate the performance of MLPs in modeling thermal transport via lattice-dynamics (LD) calculations. For clarity, we focus on the atomic forces calculated using the NEP and DP models. Figures 5(a) and 5(b) show the phonon band structures and group velocity of BAs, respectively, as calculated using the finite-displacement (FD) method. The good agreement between the dispersion curves obtained from DFT and MLPs indicates the high accuracy of our MLPs. The large acoustic-optical (*ao*) gap in BAs leads to a significant contribution of four-phonon scatterings. As depicted in Fig. 5(c), the calculated three-phonon and four-phonon scattering rates using the NEP model are in reasonable agreement with those from the DP model. Our anharmonic LD calculations based on BTE yield a $\kappa$ value of 1274.6 Wm$^{-1}$K$^{-1}$ for NEP and 1384.3 Wm$^{-1}$K$^{-1}$ for DP at 300 K, respectively. These results, along with those at other temperatures plotted in Fig. 5(d), align well with the data from our extrapolation based on MLP-MD as well as previous LD results [56]. Similar findings have also been reported in many other materials such as Si [58], $Ga_2O_3$ [22, 59], $Ba_8Ga_{16}Ge_{30}$ [58], $WS_2$ [60], and monolayer InSe [61].

To gain a deeper understanding into the performance difference between the MLP-driven MD and LD modeling of thermal transport, we calculate the root-mean-square force errors of the perturbed structures based on the NEP framework, which are used to extract the harmonic and anharmonic force constants in the FD method. The force errors are 0.19 meV/Å, 0.83 meV/Å, and 1.05 meV/Å for the second-order, third-order, and fourth-order force constants, respectively [42]. These errors are smaller by over an order of magnitude than the values obtained from the MD trajectories at 300 K based on NEP (15.46 meV/Å) and DP (14.16 meV/Å), as shown in Table S1 of the Supplemental Material [42]. This further highlights the crucial role of force errors in predicting lattice thermal conductivity with machine-learned potentials [62].

## IV. CONCLUSIONS

In summary, we systematically investigate the impact of force errors on the thermal conductivity of boron arsenide as predicted by machine-learned potentials, employing both molecular dynamics simulations (HNEMD and EMD) and lattice dynamics calculations via the phonon Boltzmann transport equation. A toy MLP model is first trained on a dataset obtained from an empirical Tersoff potential. Further, three MLP models based on the frameworks of NEP, DP, and MTP are trained based on a dataset from DFT calculations. We observe a consistent underestimation of the thermal conductivity from the MLP-MD simulations as compared to both the Tersoff-based MD simulations and previous experimental measurements. This is attributed to the effective phonon scattering due to the finite force errors in the MLPs. By introducing appropriate force noises via Langevin thermostat and adding an additional second-order term, we provide a robust extrapolation scheme to eliminate the effect of force errors and correct the underestimated thermal conductivity. Upon extrapolation to zero force error, the corrected thermal conductivity values agree well with experimental data from 200 K to 600 K. Further, predictions via the MLP-based LD calculations exhibit negligible underestimation and align closely with the corrected MD results and experimental measurements. This is because the force errors in the second, third, and fourth-order force constants are inherently small by over an order of magnitude. Our work offers deeper insight into the effect of the remaining force errors in various machine-learned potentials on thermal transport, and facilitates the prompt and accurate prediction of lattice thermal conductivity via atomistic simulations.

## DATA AVAILABILITY

The training dataset, the trained MLP models (NEP, DP, and MTP) as well as the main inputs and outputs of the MD simulations will be made freely available at [63]. The modified MLIP package with the correct heat flux is freely available at [37].

## ACKNOWLEDGEMENT

W. Z. and N. L. contributed equally. We acknowledge Penghua Ying for the helpful suggestions and discussions on the heat-flux calculations in graphene. This work was supported by the National Key R & D Project from the Ministry of Science and Technology of China (Grant No. 2022YFA1203100), the National Natural Science Foundation of China (Grant No. 52076002 and No. 12174276), and the High-performance Computing Platform of Peking University. B.S. acknowledges support from the New Cornerstone Science Foundation through the XPLORER PRIZE.

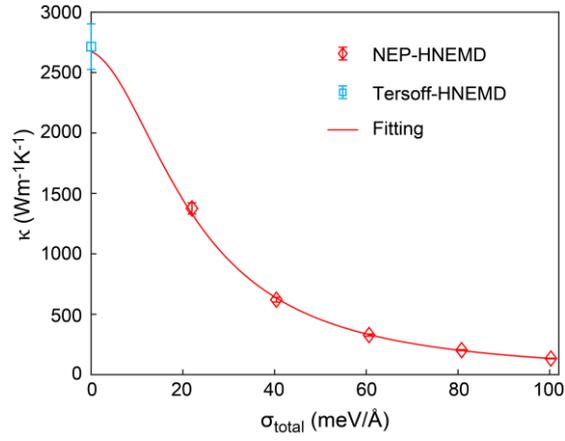

**FIG. 1. Thermal conductivity of BAs as a function of the total force error at 300 K.** The red and blue symbols are the values from HNEMD simulations based on the NEP model and the Tersoff potential, respectively. The solid line represents the fitting to the NEP-HNEMD results using Eq. (5).

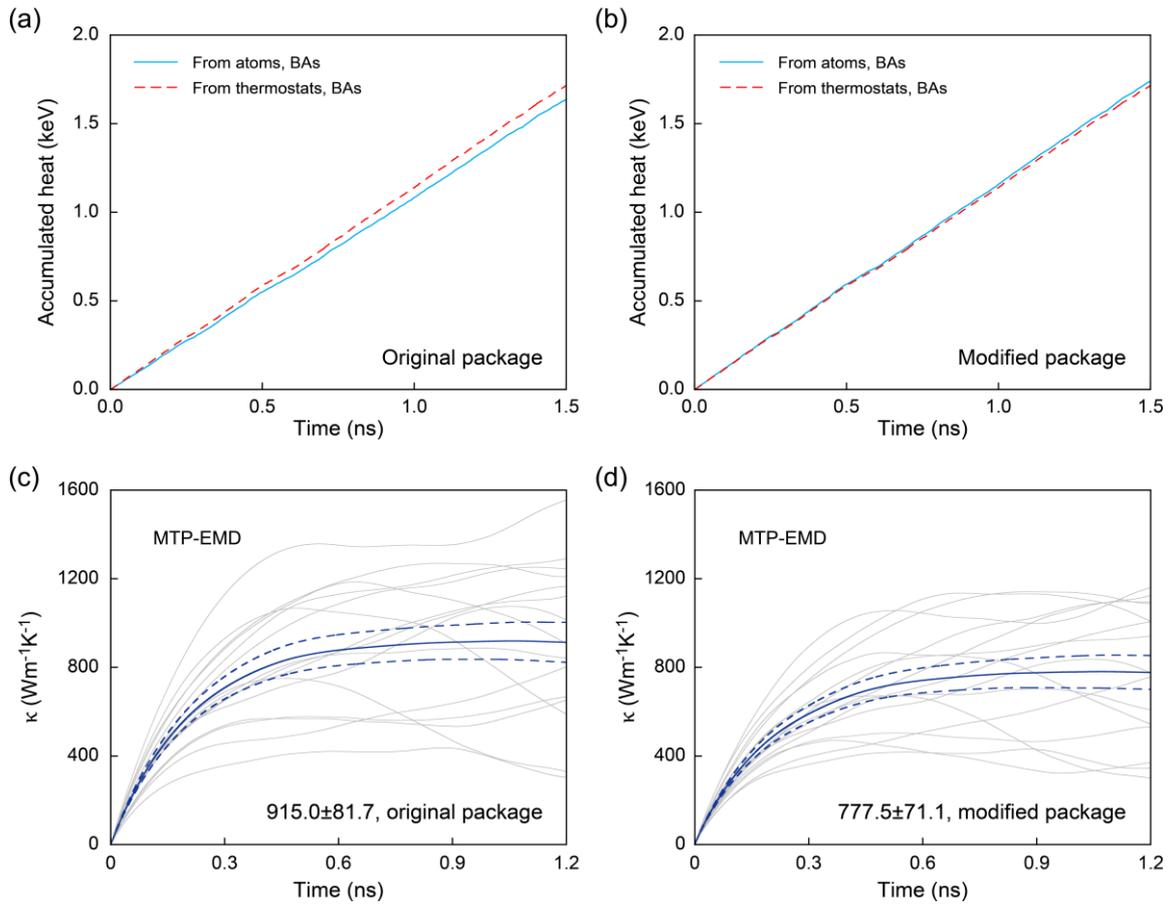

**FIG. 2. MD simulations of BAs at 300 K using the original and modified MLIP packages.** Accumulated heat in non-equilibrium steady state as a function of time using the (a) original and (b) modified MLIP package. MD simulated results averaged (blue solid line) from 15 independent runs (gray solid lines) obtained from the (c) original and (d) modified MLIP package. The dashed blue lines indicate the standard errors.

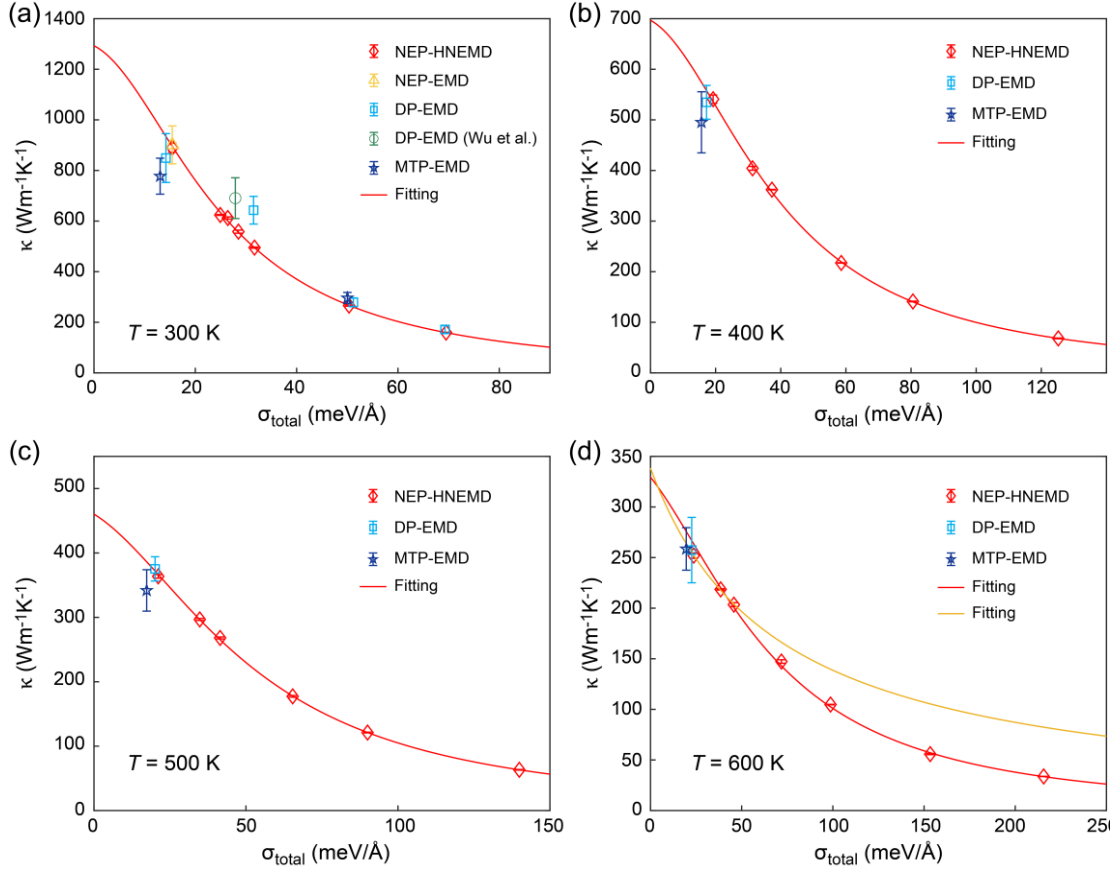

**FIG. 3. Thermal conductivity of BAs as a function of the force errors at representative temperatures.** The EMD simulated data point based on DP potential trained by Wu *et al.* [57] is plotted in panel (a) for comparison. The red solid lines represent fitting to the NEP-HNEMD simulations using Eq. (5). In panel (d), linear fitting based on Eq. (4) is also performed for comparison, using the first three data points as indicated by the orange solid line.

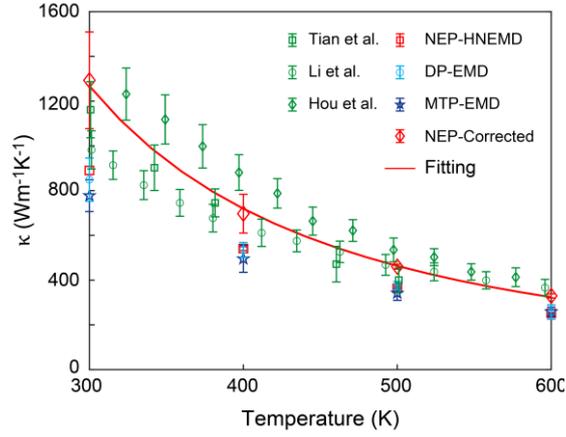

**FIG. 4. Original and corrected thermal conductivity of BAs as a function of temperatures.** The red line represents the $T^{-n}$ fitting to the corrected thermal conductivity in the zero-force-error limit from the NEP-HNEMD simulations in Fig. (3). The green symbols mark the experimental data from Refs. [31-33].

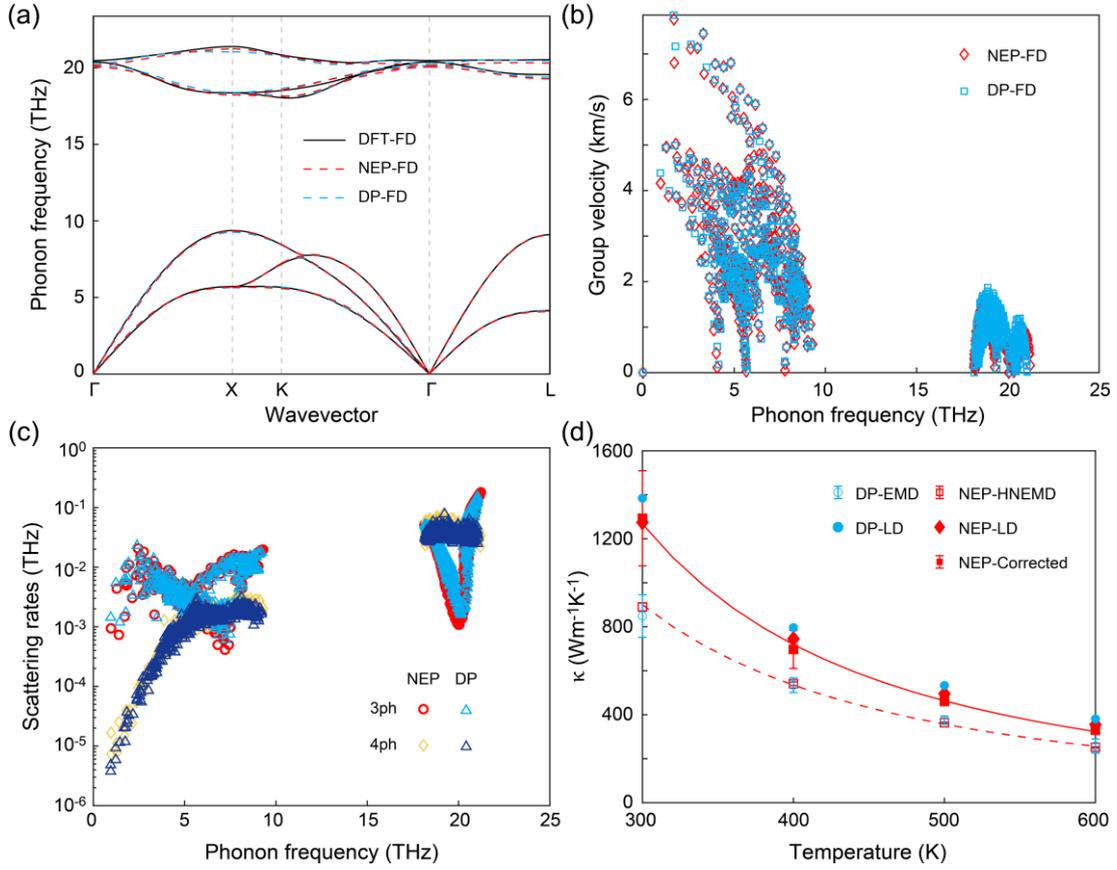

**FIG. 5. Machine-learned potential driven lattice dynamics simulations of BAs.** (a) Phonon dispersions of BAs calculated based on DFT-FD, NEP-FD, and DP-FD. (b) Group velocities of BAs from NEP-FD and DP-FD calculations. (c) Three-phonon and four-phonon scattering rates from NEP and DP calculations. (d) Calculated thermal conductivity from NEP and DP potentials as a function of temperatures. Data for the NEP-HNEMD, DP-EMD, and the NEP corrected results in Fig. 4 are also plotted for comparison.